\documentclass[mathleft
]{an}
\usepackage{graphicx}
\usepackage{times}
\newcommand{\kms}{\ensuremath{{\rm km}\,{\rm s}^{-1}}}

\newcommand{\msun}{M$_{\rm sun}$}

\begin{document}

\title{Andromeda IV, a solitary gas-rich dwarf galaxy}
\author{I.D.Karachentsev\inst{1}
\fnmsep\thanks{Corresponding author:
  \email{ikar@sao.ru}\newline}
 \and Jayaram.N.Chengalur\inst{2}
\and R.B.Tully\inst{3}
\and L.N.Makarova\inst{1}
\and M.E.Sharina\inst{1}
\and A.Begum\inst{4}
\and L.Rizzi\inst{5}}

\titlerunning{Andromeda IV, a solitary gas-rich dwarf galaxy}
\authorrunning{I.D. Karachentsev et al.}

\institute{Special Astrophysical Observatory, Nizhnij Arkhyz, 
Karachai-Cherkessia 369167, Russia
\and National Centre for Radio Astrophysics, Post Bag 3, Ganeshkhind, 
     Pune 411007, India
\and Institute for Astronomy, University of Hawaii, 2680 Woodlawn Drive, 
HI 96822, USA
\and IISER-Bhopal, ITI Campus (Gas Rahat) Building, Govindpura, Bhopal 23, India
\and W.M. Keck Observatory, 65-1120 Mamalahoa Hwy, Kamuela, HI 96743, USA 
}
\received{21 August, 2015}
\accepted{}
\publonline{later}


\abstract{
Observations are presented of the isolated dwarf irregular galaxy And IV made
 with the Hubble Space Telescope Advanced Camera for Surveys and
 the Giant Metrewave Radio Telescope in the 21 cm HI line. We 
determine the galaxy distance of $7.17\pm0.31$ Mpc using the Tip of Red Giant 
Branch method. The galaxy has a total blue absolute magnitude of --12.81 mag, 
linear Holmberg diameter of 1.88 kpc and an HI-disk extending to 8.4 times the 
optical Holmberg radius. The HI mass-to-blue luminosity ratio for And IV amounts 
$12.9~M_{\odot}/L_{\odot}$. 
From the GMRT data we derive the rotation curve for the HI and fit it
with different mass models. We find that the data are significantly
better fit with an iso-thermal dark matter halo, than by an NFW halo.
We also find that MOND rotation curve provides a very poor fit to
the data. The fact that the iso-thermal dark matter halo provides the
best fit to the data supports models in which star formation feedback
results in the formation of a dark matter core in dwarf galaxies.
The total mass-to-blue luminosity ratio of 
$162~M_{\odot}/L_{\odot}$ makes And IV among the darkest dIrr galaxies known. 
However, its baryonic-to-dark mass ratio ($M_{gas}+M^*)/M_T = 0.11$ is close 
to the average cosmic baryon fraction, 0.15.}

\maketitle

\section{Introduction}

The larger fraction of baryonic matter in spiral and elliptical galaxies 
is concentrated in stars. A typical ratio of neutral hydrogen-to-stellar
mass in spiral galaxies amounts $\sim( 3 - 10 )$\%, while in elliptical and 
lenticular galaxies the ratio $M_{HI}/M^*$ does not exceed $\sim$1\% (Roberts 
1969). In this sense, E,S0 and early-type S galaxies are situated 
at the finish line of their gas-dynamic evolution. However, there are
late-type galaxies, with still significant reserves of neutral gas remaining,
and they will be fertile over the next cosmic Hubble time, $H_0^{-1}$.

  Among about 800 galaxies in the Local Volume with distances $D < 11$ Mpc
recorded in the ``Updated Nearby Galaxy Catalog'' (= UNGC, Karachentsev et al. 
2013), 163 galaxies, i.e. 20\% of the sample, have a ratio
$M_{HI}/M^*> 1$. As it is seen from Figs. 14--19 in UNGC, the high  $M_{HI}/M^*$
ratios occur most frequently among dwarf galaxies of dIrr, dIm, and Sm types.
The galaxies with high gas-to-stellar mass ratio are usually isolated
objects, with a low optical surface brightness and low
metallicity.

  In the literature, there are instances of dwarf systems with hydrogen 
  mass-to-blue luminosity ratios of $M_{HI}/L_B \sim 10$ in solar units
(Begum et al. 2005, Warren et al. 2007, Karachentsev et al. 2008,
Chengalur \& Pustilnik, 2013). Such gas-dominated objects must be at an
initial stage of their evolution, and can be called ``laggard'' galaxies.
At the same time, the specific star formation rate of the laggard dwarfs
has a typical value of $sSFR = SFR/M^* \sim 10^{-10}~ yr^{-1}$, i.e. they are
able to reproduce their observed stellar mass during the cosmological time
$H_0^{-1}$ with their observed current star formation rate (Karachentsev \&
Kaisina 2013).

The subject of the present paper is the isolated dwarf galaxy Andromeda IV 
(= And IV) at (J2000) 00~42~32.3 +40~34~19, one of the three most HI-rich 
objects known within the Local Volume. This dwarf system of low surface 
brightness was found by van den Bergh (1972) as a candidate open stellar 
cluster at the periphery of M 31. However, deep observations of And~IV with 
the WFPC2 camera aboard Hubble Space Telescope by Ferguson et al. (2000) 
established it to be a background dIrr galaxy at a distance of 6.3$\pm$1.5 Mpc. 
Distribution of neutral hydrogen in And IV as well as the radial velocity 
field were studied by Begum et al. (2008) using the Giant Metrewave Radio 
Telescope (GMRT). The authors found that the HI- envelope around And IV 
extends to 8 times its optical Holmberg diameter, and its $M_{HI}/L_B$ ratio 
exceeds $10~M_{\odot}/L_{\odot}$. 

Below, we present  the analysis of new HST imaging and a more detailed 
analysis of archival GMRT 21 cm data of Andromeda IV.

\section{HST/ACS observations and TRGB distance}

The galaxy And IV was observed aboard HST using Advanced 
Camera for Surveys (ACS) on August 3, 2014 (SNAP 13442, PI R. Tully). 
Two exposures were made in a single orbit with the filters \textit{F606W} 
(1200 s) and \textit{F814W} (1200 s). The \textit{F606W} image of the galaxy  
And IV is shown in Fig.~1.

\begin{figure}
\includegraphics[width=3.0truein]{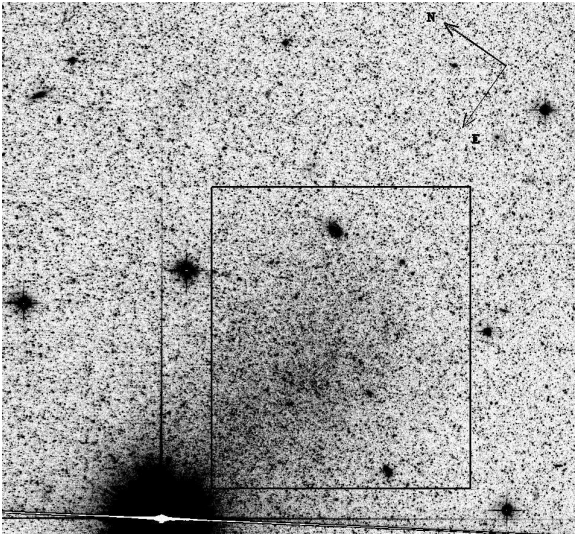}
\caption{\textit{HST}/ACS image of And~IV in \textit{F606W} filter. 
The image size is $1.6\times1.5$ arcmin.  A box of 
40\arcsec$\times$ 50\arcsec centered on And IV is shown. A box selected 
for the M31 halo population accounting is situated at the WFC1 chip about 2 
arcmin outside the And IV center.} 
\end{figure}

The photometry of resolved stars in the galaxy was performed with the ACS 
module of the \textsc{DOLPHOT} package\footnote{http://americano.dolphinsim.com/dolphot/}
for crowded field photometry (Dolphin 2002) 
using the recommended 
recipe and parameters. Only stars with photometry of good quality were 
included in the final compilation, following recommendations given in the 
\textsc{DOLPHOT} User's Guide.
 We have selected the stars with signal-to-noise (S/N) of at least 
five in the both filters. To exclude possible nonstellar objects we used in the analysis
only stars with $\vert sharp \vert \le 0.3$.

Artificial stars were inserted and recovered using the same reduction 
procedures to accurately estimate photometric errors, including crowding 
and blending effects. A large library of artificial stars was generated 
spanning the full range of observed stellar magnitudes and colours to 
assure that the distribution of the recovered photometry is adequately 
sampled.  The \textsc{DOLPHOT} package contains quite powerful and 
flexible procedure to perform artificial star experience. For an adequate 
estimation of completeness we have taken the amount of artificial stars 
10 times the number of real stars. Number of artificial stars in a particular 
part of the image depends on the distribution of real stars, and it is greater, 
the greater the real star number in the area. Thus, the completeness was 
adequately estimated in the most densely populated areas of the frame.

Figure 2 shows colour-magnitude diagrams (CMDs) for the detected stars.
The left panel presents CMD for stars within a box of 
40\arcsec$\times$ 50\arcsec centered on And IV itself. 
The contamination from M31 is extreme.  Most resolved stars are associated 
with M31 centered 42 arcmin away. The middle panel corresponds to the ACS 
field of the same area chosen outside the And IV boundary, hence samples the 
M31 halo.  The right panel reproduces a CMD for stars of And IV itself 
obtained by statistical subtraction of ``AndIV+M31'' --- ``M31'' diagrams. 
The CMDs reveal a prominent red giant branch (RGB), a red clump, and a weak 
blue plume belonging to M31 in the foreground, and a fainter RGB
attributed to the more distant And IV. The measured TRGB position for And IV
is marked by the dotted line. 

\begin{figure*}
\includegraphics[width=5.5cm]{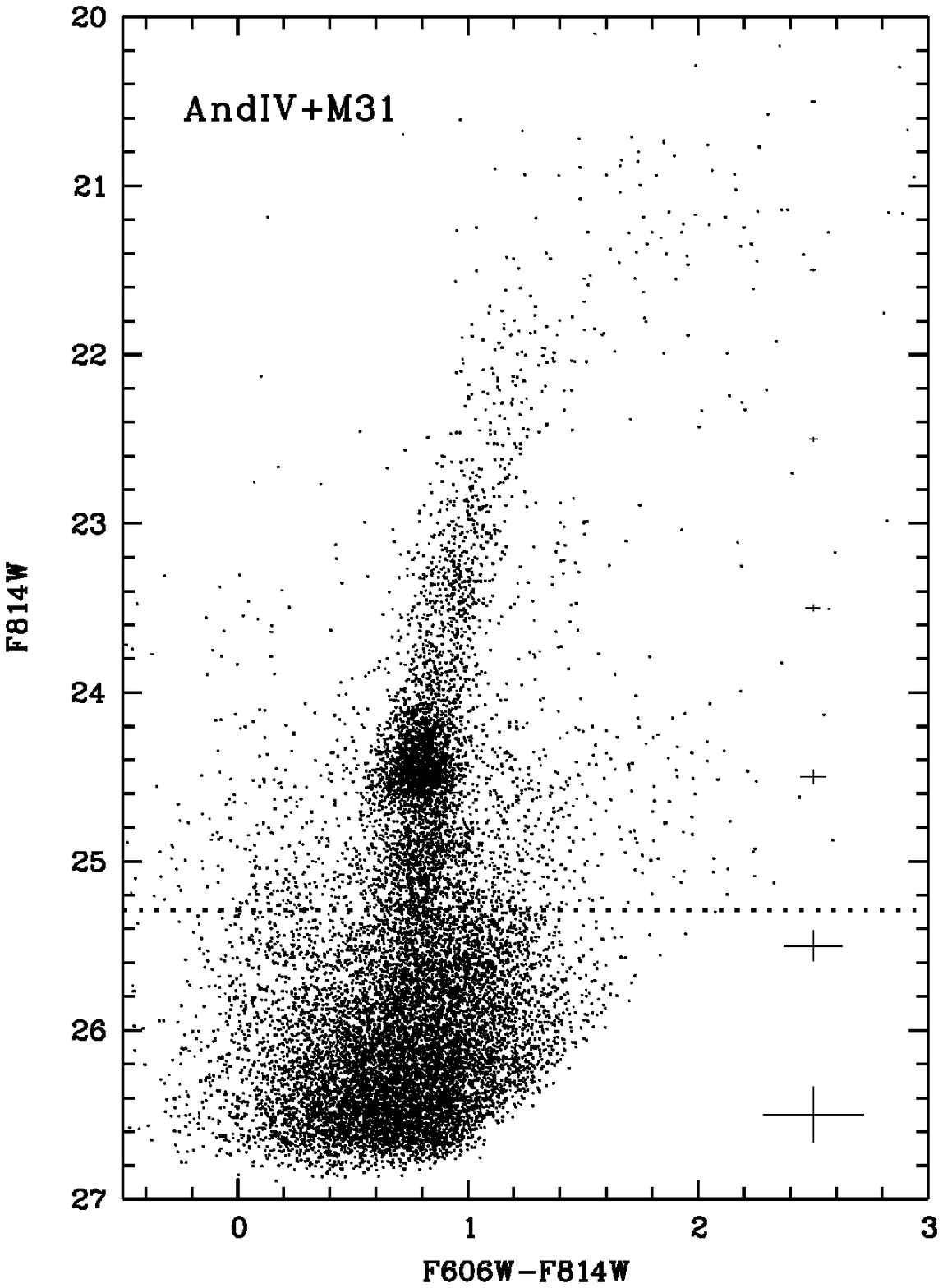}
\includegraphics[width=5.5cm]{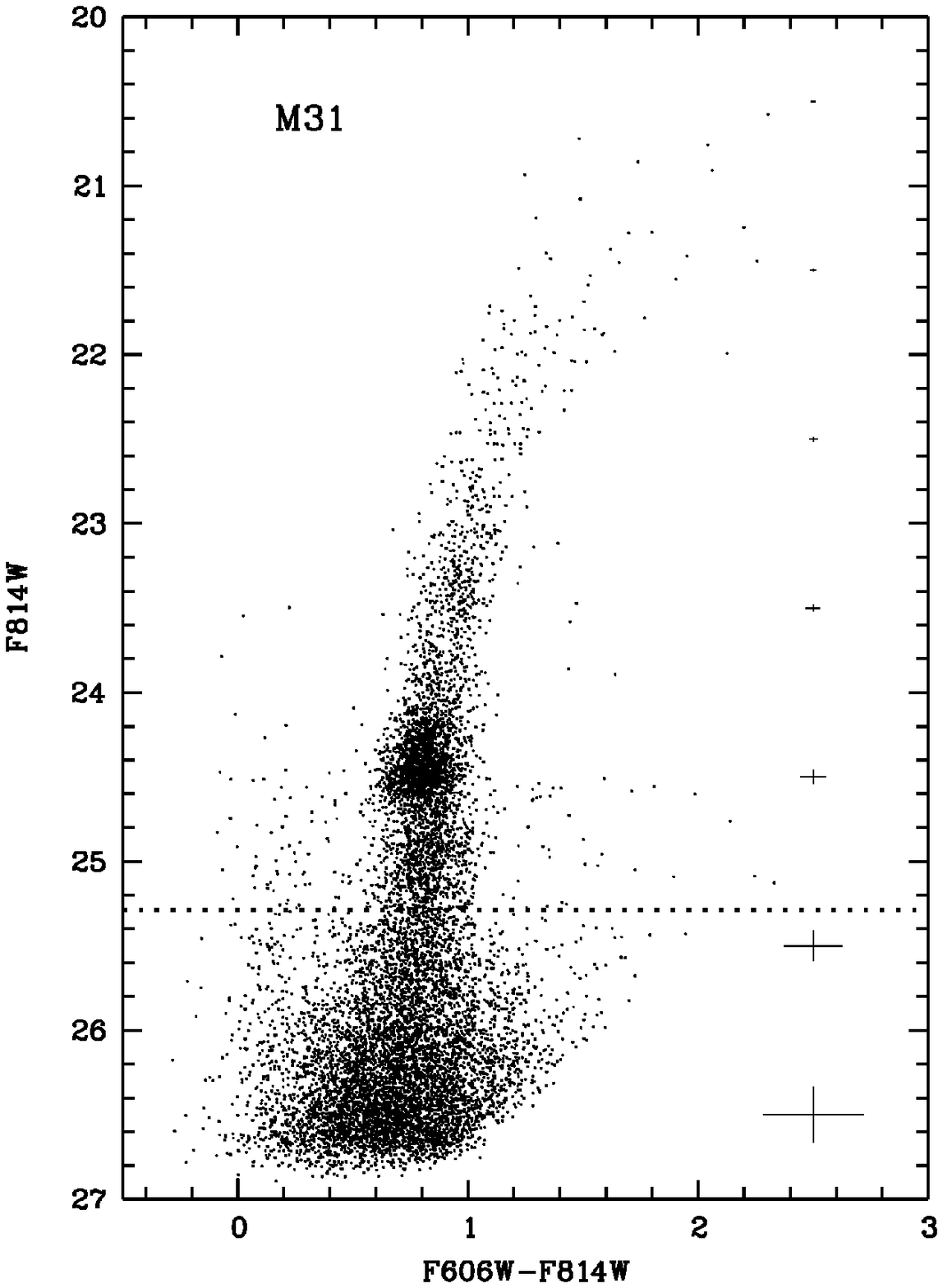}
\includegraphics[width=5.5cm]{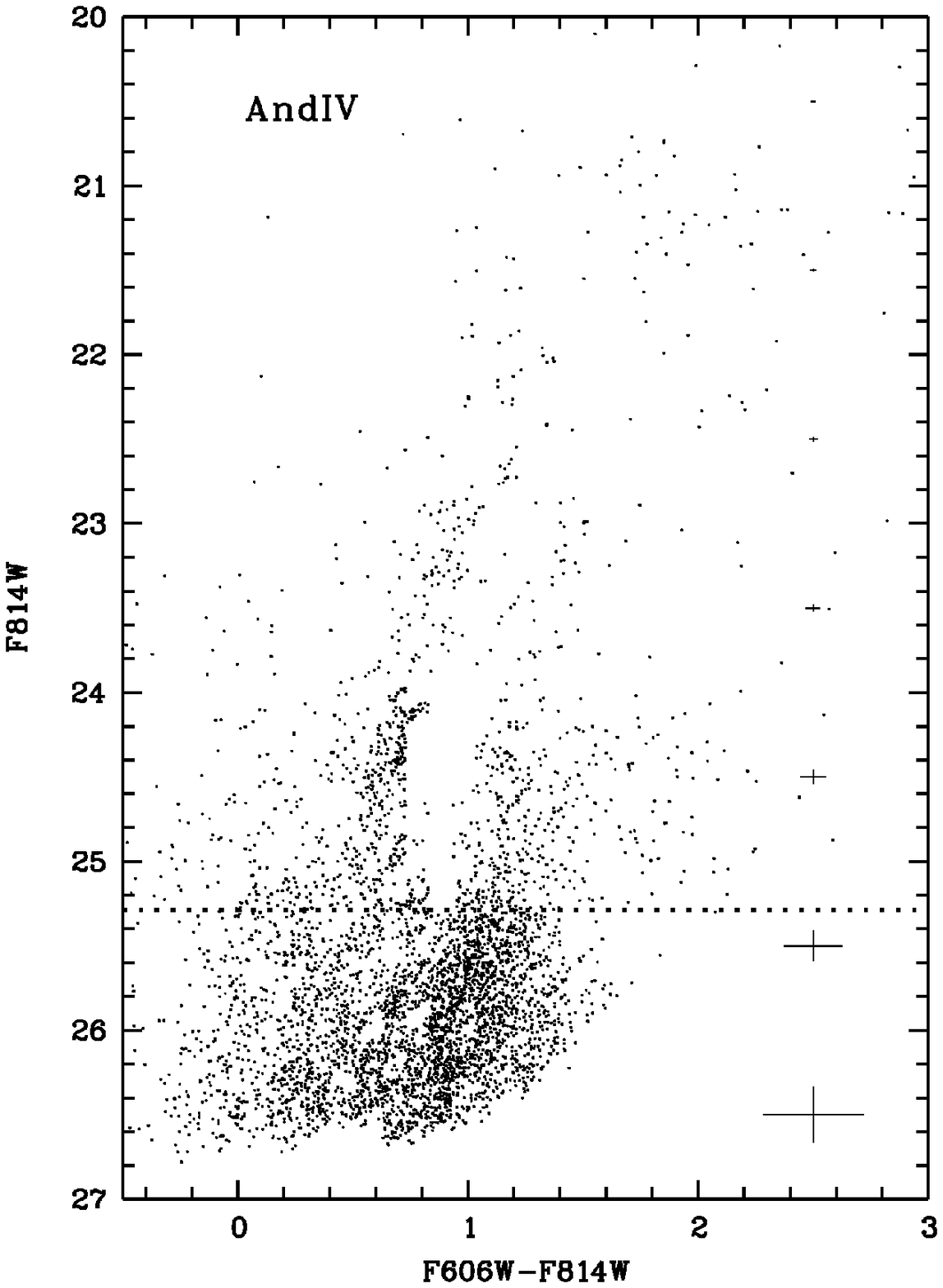}
\caption{Colour-magnitude diagrams for the detected stars in the ACS field.
The left panel presents CMD for stars within a box of 
40\arcsec$\times$ 50\arcsec centered on And~IV itself. The middle panel 
corresponds to the ACS field of the same area chosen outside the And IV 
boundary but within the M31 halo, and the right one reproduces a CMD for 
stars of And IV itself obtained by statistical subtraction of ``AndIV+M31'' ---
``M31'' diagrams.}
\end{figure*}

We have determined the And IV distance with our \textsc{trgbtool} program 
which uses a maximum-likelihood algorithm to determine the magnitude of the 
tip of the red giant branch (TRGB) from the stellar luminosity function 
(Makarov et al. 2006). The estimated value of the TRGB is 
\textit{F814W} = $25.29\pm0.07$ mag in the ACS instrumental system.
Following the calibration of the TRGB methodology developed by 
Rizzi et al. (2007), we have obtained the true distance modulus 
$(m-M)_0 = 29.28\pm0.09$ mag and the distance of $D = 7.17\pm0.31$ Mpc. 
This measurement assumes foreground reddening of $E(B-V) = 0.055$ (Schlafly 
\& Finkbeiner 2011).

\section{HST/ACS surface photometry}

Surface photometry of And IV was made with fully processed distortion-corrected 
HST/ACS \textit{F606W} and \textit{F814W} images. Some bright foreground stars 
were removed from the frames by fitting a first order surface in a rectangular 
pixel-area in the nearest neighbourhood of the star. The sky background in the ACS 
images is insignificant, but, to remove possible slight large scale variations, 
the sky was approximated by a tilted plane, created from a two-dimension 
polynomial, using the least-squares method. The accuracy of the sky background 
determination is about 2\% of the original sky level. The main source of the 
background uncertainty was the bright red star east of And IV seen in Fig.~1.

To measure total galaxy magnitude in each band, the galaxy image was first 
fitted with concentric ellipses. Then integrated photometry was performed in 
these ellipses with parameters defined from the centre to the
faint outskirts (Bender \& Moellenhoff 1987). The total 
magnitude was then estimated as 
the asymptotic value of the radial growth curve. The measured total 
magnitude and colour are equal to $V = 16.2\pm0.15^m$ and $(V-I) = 0.75\pm0.05$. 
The estimated errors include the photometry and sky background uncertainties,
as well as the transformation errors from instrumental ACS magnitudes to the 
standard $V$ and $I$ magnitudes (Sirianni et al. 2005).

 Azimuthally averaged surface brightness profiles for And IV were obtained by 
differentiating the galaxy growth curves with respect to semiaxes. The 
resulting profiles in $I$ (upper line) and $V$ bands are displayed in Fig. 3. The low 
surface brightness of the galaxy makes the profiles rather noisy. 

\begin{figure}
\includegraphics[width=5.6cm,angle=-90]{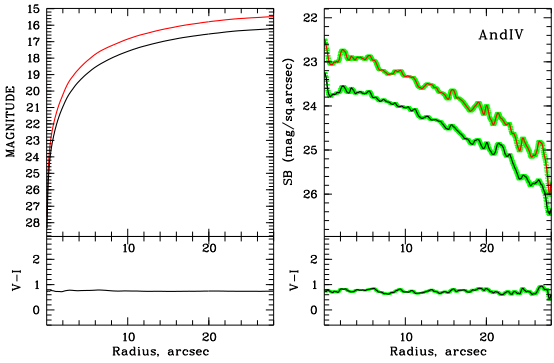}
\caption{Azimuthally averaged brightness profiles and $V-I$ colour
    for And~IV obtained from the HST/ACS field,  where 1 arcsec 
    corresponds to 34.8 pc. The left panel shows
    the cumulative magnitude and  colour while the right panel
    shows the surface brightness and colour. A negligible colour 
    gradient is seen along the galaxy radius.}
\end{figure}

 As known, surface brightness profiles of dwarf irregular and spheroidal
galaxies can be fitted by an exponential intensity law of brightness 
distribution in magnitudes per square arcsec (de Vaucouleurs 1959):

    $$\mu(r) = \mu_0 + 1.086 \times (r/h),$$
where $\mu_0$ is the central surface brightness and $h$ is the exponential 
scale length. The surface brightness profiles of And~IV are moderately well 
fitted by an exponential law. The unweighted exponential fits to the profiles 
yield the central surface brightness of $\mu_0^V = 23.6\pm0.1$ and 
$\mu_0^I = 22.8\pm0.1$. The scale lengths are $h^V = 12.6\arcsec$ and 
$h^I = 13.7\arcsec$.

Some basic photometric parameters of And~IV are given in Table 1, where 
the transformation from $V$ band to $B$ one was done using the standard 
relation

 $$B - V = 0.85 (V - I) -0.2 $$
valid for dIrr galaxies (Makarova, 1999).


\section{GMRT HI- observations and data analysys}

The moment maps for And~IV have been previously published in Begum et al. 
(2008). We summarize briefly below the observational and data reduction
details, and refer the reader to the earlier paper for more details.
The GMRT (Swarup et al. 1991) observations of And~IV  were conducted on
2$^{nd}$ Jan 2007. An observing bandwidth of 1 MHz centered at 1418.65~MHz (which
corresponds to a heliocentric velocity of $\sim 235$~\kms) was used . The band
was divided into 128 spectral channels, giving a channel spacing of 1.65 \kms.
Absolute flux calibration was done using scans on the standard calibrator
3C48, which were observed at the start and end of the observing run.
Phase calibration was done using 0029+349, which was observed once every 50
minutes. Bandpass calibration was done in the standard way using the 
3C48 observations. The total on-source time was $\sim$ 8.6 hours. Data
analysis was done using the classic AIPS package, and data cubes made
at spatial resolutions of $\sim 41^{''},24^{''}$ and $11^{''}$.

\begin{figure}
\includegraphics[width=3.0truein]{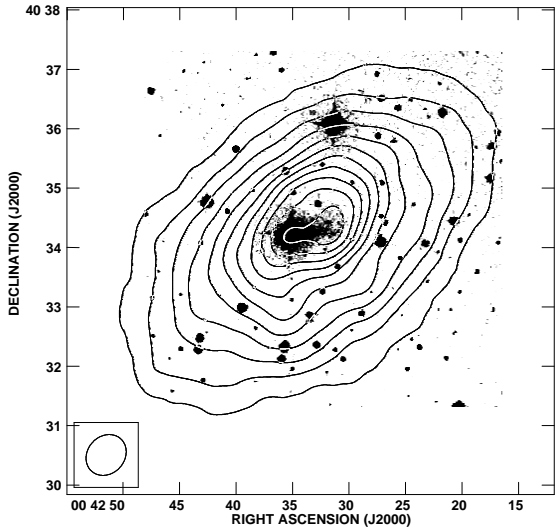}
\caption{The integrated HI column density distribution
of And~IV at $44^{''}\times 38^{''}$ resolution overlayed on optical DSS image. 
}
\label{fig:mom0}
\end{figure}

Fig.~\ref{fig:mom0} shows the integrated HI emission from And~IV at 
44$''\times 38''$ resolution, overlayed on the digitised sky survey (DSS) 
image. As can be seen, the HI distribution of And~IV is regular; the total 
HI extent at a column density of 5$\times 10^{19}$ cm$^{-2}$ is 7.2$^{\prime}$.
The total extent at a column density of  1$\times 10^{19}$ cm$^{-2}$ is 
7.6$^{\prime}$ (or $\sim 8.4$ times the Holmberg diameter). The integrated HI 
flux obtained from this data is 17.4 Jy~km/s, which is slightly less than 
the value of 22.4 Jy~km/s reported by Braun et al. (2003). 

\begin{figure}
\includegraphics[width=3.0truein]{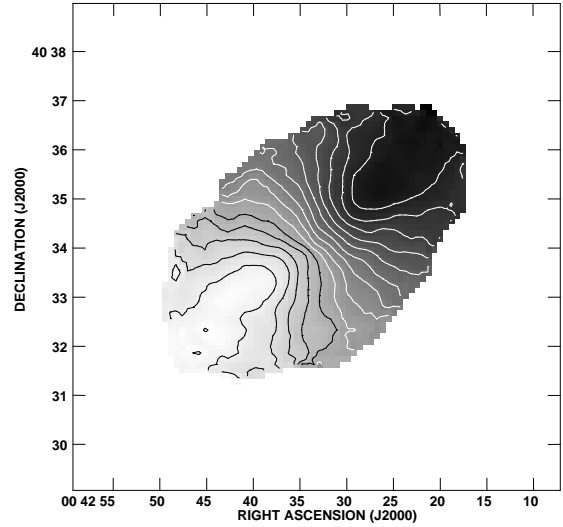}
\caption{The HI velocity
        field of And~IV at 26$^{''}\times 23^{''}$ resolution. The
        contours are in the steps of 5 \kms and range from 180.0~\kms
        to 265.0~\kms.
}
\label{fig:mom1}
\end{figure}

Fig.~\ref{fig:mom1} shows the velocity field of And~IV at 26$^{''}\times 23^{''}$ 
resolution. The velocity field is regular and a large scale velocity gradient,
consistent with systematic rotation, is seen across the galaxy. The velocity 
field of this galaxy is lopsided; the isovelocity contours in the south-eastern
half of the galaxy are more closed than the north-western half.  However, on
the whole, in comparison with the  galaxies in the FIGGS survey (Begum et al
2008) the velocity field for And~IV is one of the most symmetric and
undisturbed velocity fields of a faint dwarf galaxy.

\subsection{HI rotation curve}
\label{ssec:rotcur}

\begin{figure}
\includegraphics[width=3.0truein]{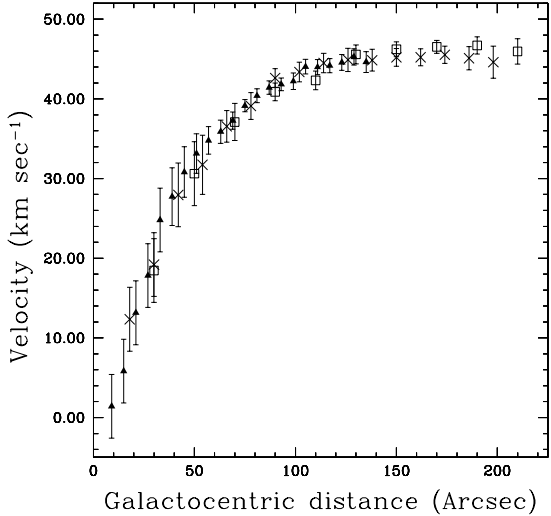}
\caption{ 
The rotation curves derived from the intensity
weighted velocity field at various resolutions.
Triangles, crosses and squares show the rotation velocity derived from
 12$^{''}\times11^{''}$,26$^{''}\times23^{''}$,
and 44$^{''}\times38^{''}$ resolution  respectively.
}
\label{fig:vrot}
\end{figure}

\begin{figure}
\includegraphics[width=3.0truein]{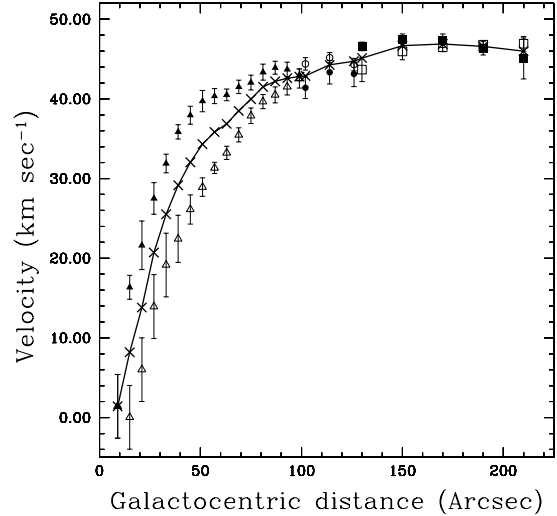}
\caption{ 
Filled and open points show the hybrid rotation curve derived separately
from the receding and the approaching side respectively for various resolution 
viz. triangles (12$^{''}\times11^{''}$), circles (26$^{''}\times23^{''}$)
and squares(44$^{''}\times38^{''}$). The adopted hybrid rotation curve 
(crosses) is an average of the rotation velocities derived from the 
approaching and receding halves. 
}
\label{fig:vrot_allsides}
\end{figure}

   Rotation curves of And~IV were derived using 44$^{''}\times 38^{''}$, 
26$^{''}\times 23^{''}$ and 12$^{''}\times 11^{''}$ resolution velocity fields, 
using tilted ring fits. The center and systemic velocity for the galaxy 
obtained from a global fit to the various resolution velocity fields matched 
within the error bars; the best fit systemic velocity of 234.0$\pm$1.0 \kms 
matches  well with the value obtained from the global HI profile of the galaxy.
Keeping the center and systemic velocity fixed, we fitted for the inclination 
and position angle (PA) in each ring. For all resolution velocity fields, the 
PA was found to vary from $\sim 307^\circ~\rm{to}~314^\circ$ and
the inclination varied from $\sim 65^\circ~\rm{to}~50^\circ$.
Keeping the PA  fixed to 310$^\circ$ at all radii and  inclination  fixed 
to 62$^\circ$ in the inner regions (upto $\sim$ 100$^{''}$) and 52$^\circ$ in the
outer regions, the rotation curves at various resolutions were derived. Fig.
~\ref{fig:vrot} shows the rotation curve of the galaxy derived at various 
resolutions $-$ the rotation curves at various resolutions match within the 
errorbars.  A hybrid rotation curve for And~IV is obtained by using the 
12$^{''}\times 11^{''}$ resolution velocity field in the inner regions and the 
low resolution velocity fields in the outer regions. The rotation curves were
also derived at different resolutions using the approaching and receding 
halves of the galaxy separately. The rotation curve in the inner regions 
of the galaxy, derived for each half, was found to be somewhat different 
from the curve derived using the whole galaxy (Fig.~\ref{fig:vrot_allsides}). 
For the purpose of mass modelling we have used a mean of
the rotation curves for the two sides. The adopted hybrid mean rotation curve 
is shown as a solid line in Fig.~\ref{fig:vrot_allsides}. The errorbars on 
the mean rotation curve were obtained by adding  quadratically  the 
uncertainty reported by  the tilted-ring fit as well as the difference in 
rotation velocities between the approaching and receding side. The 
"asymmetric drift correction" (see e.g. Begum et al. (2003) for a discussion
of the asymmetric drift correction to rotation curves of dwarf galaxies) was 
found to be  small compared to the errorbars at all radii and hence ignored. 
The rotation curve shows a flattening beyond $\sim$ 150$^{\prime\prime}$; 
And~IV is one of the faintest known dwarf galaxy to show a clear flattening 
of the rotation curve.

\subsection{Mass Models}

\begin{figure}
\includegraphics[width=6cm,angle=-90]{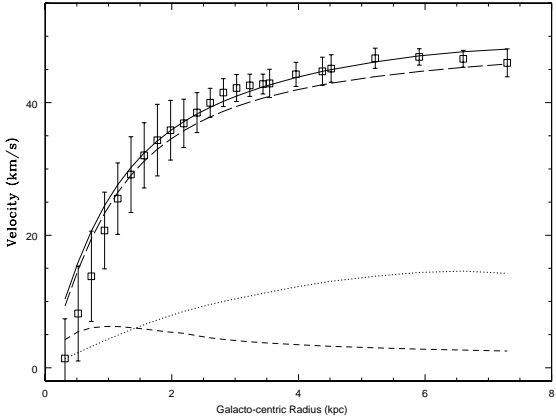}
\caption{Iso-thermal halo based mass model for AndIV using the mean hybrid 
  rotation curve (points with error bars. The dashed line shows the 
  contribution of the stellar disk to the total rotational velocity, the 
  dotted line shows the contribution of the gas disk, while the long-dashed 
  line shows the contribution of the dark matter halo. The solid line shows the 
  quadrature sum of all of these individual components. See the text for 
  more details.
}
\label{fig:iso}
\end{figure}

\begin{figure}
\includegraphics[width=6cm,angle=-90]{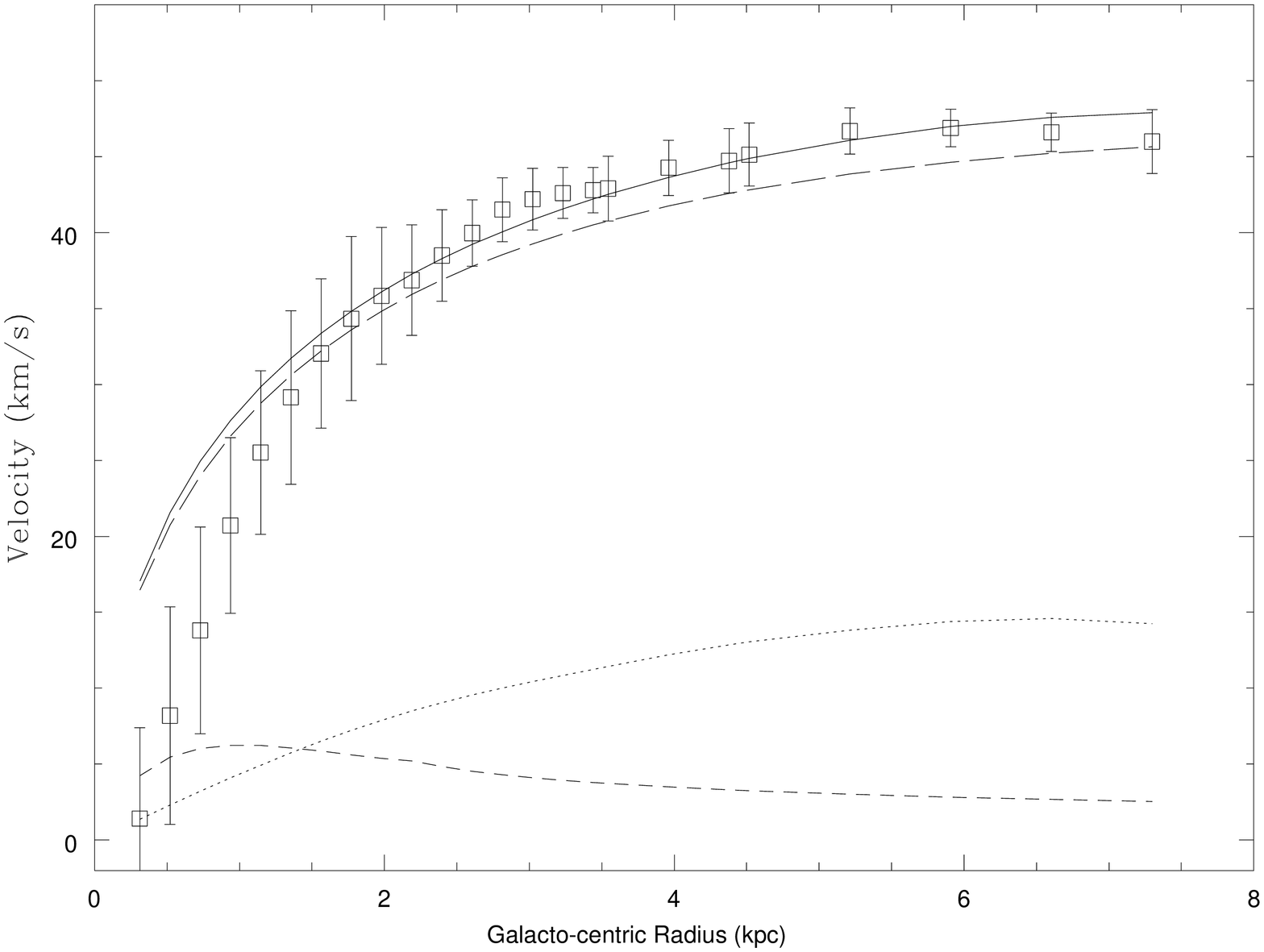}
\caption{NFW halo based mass model for AndIV using the mean hybrid 
  rotation curve (points with error bars. The dashed line shows the 
  contribution of the stellar disk to the total rotational velocity, the 
  dotted line shows the contribution of the gas disk, while the long-dashed 
  line  shows the contribution of the dark matter halo. The solid line shows 
  the  quadrature sum of all of these individual components. See the text 
  for more details.
}
\label{fig:nfw}
\end{figure}

Mass models for the galaxy were constructed using the GIPSY task rotmas.
The galaxy was modelled as consisting of 3 components, a stellar disk,
a gas disk and dark matter halo. The structural parameters of the stellar disk 
were fixed using the photometry in I-band presented in this paper. The remaining 
parameter required for mass modelling is the mass to light ratio of the
stellar disk. Leaving this as a free parameter in the fit lead to unphysical
results. We hence fix it to a value of 0.7, which is reasonable given the
 colors of the galaxy (The M/L ratio derived from the B-V color and from the 
 Bell \& de Jong (2001) model with low metallicity (Z=0.02), 
Bruzual $\&$ Charlot SPS model using a modified Salpeter IMF, is 0.74).
In any case, the stellar disk is dynamically not dominant, and varying 
the mass to light ratio by a factor of 2 changes the deduced parameters 
of the dark matter halo by $\sim 5\%$.  The rotation
velocity of the gas disk was determined from the measured surface density
of the disk. The HI surface density was multiplied by a factor of 1.3 to
account for the presense of primordial helium. The molecular gas content
of dwarf galaxies is expected to be small (see e.g. Taylor et al. 1998; 
Cormier et al. 2014) and we hence neglect it in the mass modelling.
Two different mass models were constructed, one which assumed that the halo
profile is that of a modified iso-thermal halo, and the other which assumes
that the halo has an NFW type profile. The two fits are shown in 
Figs.~\ref{fig:iso} and ~\ref{fig:nfw}. The reduced $\chi^2$ for the 
iso-thermal profile fit (0.37) is more than a factor of two smaller than 
that for the NFW halo. For the iso-thermal halo the best fit core density and
core radius are 52.4~\msun/pc$^{-2}$ and 0.953~kpc, while for the NFW
halo the best fit gives a concentration parameter $c= 5.84$ and 
R$_{200} = 32.3$~kpc. From the scaling relations for LCDM halos (see e.g.
Bullock 2001; Bottema 2015) the expected concentration parameter is 

\begin{equation}
c = 55.74 ( {V_{max} \over \kms } )^{-0.2933}
\end{equation}

From the rotation curve the maximum rotation velocity is $\sim 45$~\kms
which gives the expected value of $c = 18.25$, significantly different
from what we get from the best fit. In fact a concentration parameter
$\sim 18$ is unable to give a reasonable fit the rotation curve for 
any value of R$_{200}$. We also note that the fact that the iso-thermal
halo provides the best fit to the rotation curve supports models in which 
feed back from star formation leads to the formation of cores in the 
centres of dwarf galaxies (e.g. Governato et al. 2010).  Since the
biggest difference between the fit of the isothermal and the NFW profile
usually seen in the galaxy center, we analyzed also the effect of
the HI data resolution on this difference as well as the effect of
a separate fit for the approaching and receeding sides. Both the effects
are found to be negligible to explane the discussed difference.

\begin{figure}
\includegraphics[width=6cm,angle=-90]{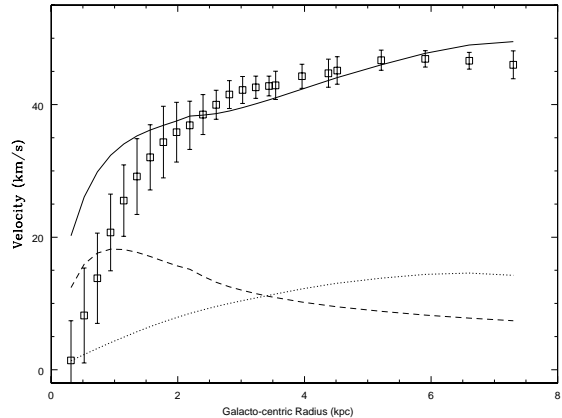}
\caption{MOND based mass model for AndIV using the mean hybrid 
  rotation curve (points with error bars. The dashed line shows the 
  contribution of the stellar disk to the total rotational velocity, the 
  dotted line shows the contribution of the gas disk. The solid line shows the 
  expected rotation curve under MOND. See the text for more details.
}
\label{fig:mond}
\end{figure}

Finally, in Fig.~\ref{fig:mond} we show the expected rotation curve
under the MOND hypothesis (Milgrom, 1983). As can be seen, the fit
is quite poor, even though the Mass to Light ratio of the stellar
disk has been left as a free parameter in the fit. The best fit
values of the mass to light ratio are $5.9 \pm 2.0$ (in solar units),
and the best fit MOND a$_0$ parameter is 
$(1.03 \pm 0.15) \times 10^{-8}$~cm~s$^{-2}$. Although the fit quality
is very poor compared to that for the Isothermal and NFW halos (the
reduced $\chi^2$ is 2.2), we note that the best fit value for a$_0$
agrees within error bars with the fiducial value of 
$1.21 \times 10^{-8}$~cm~s$^{-2}$ (Begeman, Broeils \& Sanders, 1991).

From the rotation curve, the total dynamical mass of And~IV is 
$3.4 \pm 0.3)\times 10^9$\msun. The total hydrogen mass is given by
$M_{HI} = 2.356\times 10^5 D^2 F(HI),$ where $D$ in Mpc and the total galaxy 
flux in Jy km/s ( Roberts \& Haynes, 1994). Using the measurement 
of $22.4\pm0.5$ Jy km/s by Braun et al. (2003) we derive the hydrogen 
mass to be $2.7\times10^8$~\msun. The total gas
mass (including the mass of helium) is $\sim 3.5 \times 10^8$~\msun. 
This quantity is more than one order larger than the total stellar mass 
of And~IV, viz. $M_{*} = 0.27\times10^8$ (UNGC). Within the last measured
point of the rotation curve, the ratio of the baryonic mass to the
total mass is $\sim 0.11$. With the new estimates of apparent blue magnitude 
and the distance, the absolute magnitude of And~IV equals to $M_B = -12.91$ 
that yields the hydrogen mass-to-blue luminosity ratio $M(HI)/L_B = 12.9$ and 
the total mass-to-luminosity ratio $M_T/L_B = 162$. Both quantities are among 
the highest for any known galaxy. The basic observed and distance-dependent 
properties of And~IV are summarized in Table 1.

One can ask whether the extremely gas rich dwarf galaxies have abnormally small 
baryon fractions, i.e. have they just been inefficient at forming stars, or
did they end up with less than the typical baryon fraction? The ratio of
baryonic-to-dark  matter is expected to systematically vary with halo mass, 
since small halos are both inefficient at capturing hot baryons during the 
epoch of reionization and also because small halos easely loose baryons as
a consequence of energy input from star formation.
Taking into account a correction for abundance of helium and molecular gas
 we obtain the baryon fraction
(as determined at the last measured point of the rotation curve) to be
$f_b = (M_{gas} + M^*)/ M_T = 0.11$. As can be seen, the baryonic fraction 
in And~IV is fairly close to the average cosmic baryon fraction 
$f_b$(cosmic) = 0.15 (Planck collaboration, 2015). As such, the And~IV has
got its ``fair share'' of  baryons, but for some reason has been unable to 
convert them into stars.

\section{Spatial environment and SFR.}   

The dwarf system And IV resides in a region of low spatial density by number 
of galaxies. Its nearest neighbors are the Sc galaxy NGC 7640 with
absolute magnitude $M_B = -19.2$ at a 3D-separation of 2.0 Mpc, and the pair of
galaxies NGC 672/ IC 1727 with $M_B = -18.8^m/-17.8^m$ separated by 2.2 Mpc.
The mean density of stellar matter in a sphere of 1 Mpc radius around And IV
consists of $\sim$1/25 of the mean cosmic stellar density, which indicates the 
severe isolation of And IV.
  
Observations in the $H_{\alpha}$ line by Ferguson et al. (2000) and Kaisin \&
Karachentsev (2006) found five emission knots, with integral $H_{\alpha}$ flux
corresponding to a star formation rate of 
log $SFR(_{H\alpha}) = -3.12 (M_{\odot}$/yr). The estimate of $SFR$ via a flux 
in the far ultraviolet (Gil de Paz et al. 2007, Lee et al. 2011) is 5 times greater: 
$\log SFR(FUV) = -2.42 (M_{\odot}$/yr) that is typical for dIrr of low mass
(Lee et al. 2011, Fumagalli et al. 2011, Weisz et al. 2012).  This difference 
is partially caused by stars of different age (temperature), depending on the
details of starformation history on scales of 10 and 100 Myrs (Calzetti 2013).
The corresponding estimates of the specific star formation rates are 
$\log sSFR(H\alpha) = -10.56$ and $\log SFR(FUV) = -9.86 $/yr) being 
comparable with the value of Hubble parameter, $\log H_0 = -10.14$ /yr.

Pustilnik et al (2008) carried out spectral observations of two HII-regions
in And~IV and found their metallicity to be $\sim$1/14 the 
solar value. The presence of a significant amount of neutral hydrogen, a 
moderate activity of star formation and a low metallicity are typical features 
of isolated dwarf irregular galaxies (Huang et al. 2012, 
Karachentsev \& Kaisina 2013, Cannon et al. 2015).

  \section{And IV among other gas-rich galaxies in the Local Volume.}

At present, the UNGC catalog is the most representative sample of galaxies 
limited by galaxy distance, but extending to faint apparent magnitudes and 
HI-flux. Dwarf galaxies, in particular gas-rich systems, are the major 
constituent of the UNGC. Among eight hundred UNGC galaxies we selected
16 that satisfy the condition: 
     $$B^c - m_{21} > 1.0, $$
where $B^c$ is the apparent blue magnitude of a galaxy corrected for Galactic
extinction, and  $m_{21} = 17.4 - 2.5 \log F(HI)$ is its 21-cm line flux in 
magnitudes as defined in HyperLeda (Paturel et al. 1996). Assuming the stellar
mass of a galaxy expressed via its luminosity in the K-band is $M^*/L_K =
1.0 M_{\odot}/L_{\odot}$ (Bell et al. 2003), we derive
  $$\log (M_{HI}/M^*) = 1.02 + 0.4 (m_K - m_{21}).$$
For the dwarf galaxies of dIrr, dIm types there is a relation
$\langle B^c - m_K\rangle = 2.35$ (Jarrett et al. 2003). Therefore, for dIrr
galaxies, the condition  $B^c - m_{21} > 1.0$ corresponds to the ratio:
   $$\log( M_{HI}/M^*) = 0.08 + 0.4( B^c - m_{21}) > 0.48.$$
In other words, the hydrogen mass of these galaxies exceeds their stellar
mass by at least a factor three.

Basic observables of the 16 galaxies selected from UNGC and ranked by
the $(B^c - m_{21})$ value are given in Table 2.
Its columns contain: (1) galaxy name, (2) apparent blue magnitude, (3)
Galactic extinction in the B-band (Schlafly \& Finkbeiner, 2011), (4) the
mean surface brightness in the B-band within the Holmberg radius, (5)
radial velocity (km/s) in the Local Group rest frame, (6) the  21-cm line
flux in magnitudes, (7) HI line width (km/s), measured at the 50\% level from
the maximum, (8) distance in Mpc, (9) stellar mass in solar units, (10)
hydrogen mass in solar units, (11) the tidal index, used in UNGC, with
negative values corresponding to isolated galaxies, (12) hydrogen-to-stellar
mass ratio, (13) difference of the optical and 21-cm magnitudes used for 
the object selection. We excluded from this list 4 dwarf galaxies: BK3N,
Holmberg IX, d0226+3325 and DDO 169NW, having $B^c - m_{21} > 1.0$ but 
locating in contact with brighter neighbors, because their HI-flux is 
difficult to separate from the HI-dominant galaxy.

It should be mentioned, that apparent magnitudes of some faint galaxies
(for instance, AGC208387, AGC208399) have been estimated visually with a
typical error of $\sim0.4^m$. Note also, that the present list of HI-richest
galaxies in the Local Volume may be still incomplete.

The last line in Table 2 presents the mean observables for the 16 objects 
that are discussed. Gas-rich galaxies are dwarf systems
with a characteristic stellar mass of $M^*/M_{\odot}$ = 7.14 dex, HI-line 
width of 45 km/s, a low surface brightness of 25.5$^m/\sq\arcsec$, and the 
HI-mass-to-stellar mass ratio of 5.0. Judging by the mean value of the tidal 
index, $\langle\Theta_1\rangle = -1.0$, the majority are well isolated 
galaxies. The isolated dIrr galaxy And IV is one of the three most gas-rich 
galaxies in the Local Volume. Some other galaxies from Table 2: ESO215-009, 
KK144, DDO143, KK195, have been already studied in HI with high angular 
resolution (Warren et al. 2004, Begum et al. 2008). It will be interesting 
to obtain detailed maps of HI-density and velocity fields for other objects 
of the sample.

\acknowledgements

The Space Telescope Science Institute provided support in connection with
observations on Hubble Space Telescope with program GO--13442.  
 We thank the staff of the GMRT who made these observations possible.
   The GMRT is operated by the National Center for Radio Astrophysics of
   the Tata Institute of Fundamental Research.
The work in
Russia is supported by the grant of the
Russian Scientific Foundation 14--12--00965 and the RFBR grant 15--52--45004.
LNM acknowledge the support from RFBR grant 13--02--00780 and Research 
Program OFN of the Division of Physics, Russian Academy of Sciences.

{}

\begin{table*}
 \centering
 \caption{Global parameters of AND~IV.}
 \begin{tabular}{lcl}\hline
 Parameter                     &  Value (error)  &   Reference\\
\hline
 R.A.,Dec.(J2000.0)            &004232.3+403418  &    van den Bergh S., 1972\\
 Galactic extinction in B, mag &      0.22       &    Schlafly \& Finkbeiner, 2011\\
 Internal extinction in B, mag &      0.04       &    UNGC\\
 Morphological type            &      Ir--LSB    &    UNGC\\
 Tidal Index, $\Theta_1$          &      --1.9   &    UNGC\\
 $B_T$, mag                     &   16.64 (0.10) &    this study\\
 $(B-V)_T$                     &    0.46 (0.05)  &    this study\\
 $\mu_0^B$,  mag arcsec$^{-2}$   &   24.1 (0.1)  &    this study\\
 Holmberg diameter, arcmin    &    0.90 (0.03)   &    this study\\
 Scale length, arcsec         &      13 (1)      &    this study\\
 $V_{hel}$,  km/s                 &     234 (1)  &    this study \\        
 $W_{50}$,  km/s                  &      90 (2)  &    Begum et.al, 2008\\
 S(HI), Jy km/s               &    22.4 (0.5)    &    Braun et.al, 2003\\
 HI-diameter, arcmin          &     7.6 (0.5)    &    Begum et.al, 2008\\
 HI-axial ratio               &    0.61 (0.02)   &    this study\\
 $B_0 - m_{21}$, mag             &    2.38 (0.15)&    this study\\
 m(FUV)                      &   17.90 (0.10)    &    Lee et al, 2011\\
 $H_{\alpha}$ flux, erg/cm$^2$ sec &  --13.88 (0.07)& Kaisin \& Karachentsev, 2006\\
 Distance,  Mpc              &    7.17 (0.31)    &    this study \\           
 $M_B$,  mag                   &  -12.91 (0.10)  &    this study \\
 Holmberg diameter, kpc      &    1.88 (0.06)    &    this study \\
 $\log(M^*/M_{\odot})$              &       7.44 &    UNGC \\
 $\log(M_{HI}/M_{\odot})$           &       8.43 &    Braun et.al, 2003\\
 $\log(M_T/M_{\odot})$              &       9.53 &    this study\\
 $\log(SFR), H_{\alpha}$       &   --3.12 (0.07) &    Kaisin \& Karachentsev, 2006\\
 $\log(SFR), FUV$          &   --2.42 (0.04)     &    Lee et al, 2011\\
\hline
\end{tabular}
\end{table*} 

\begin{table*}
\centering
 \caption{Properties of 16  LV galaxies with $B_c-m_{21} > 1.0^m$}
 \begin{tabular}{llcccccccrrrc}\hline
Galaxy       &   $B$  &  $A_b$ &  SB &$V_{LG}$ & $m_{21}$& $W_{50}$&   $D$  &  $\log M^*$& $\log M_{HI}$&   $\Theta_1$& 
$\log M_{HI}/M^*$& $B_c-m_{21}$\\
\hline
ESO215-009   & 16.0 & 0.95& 25.6& 290 & 12.4&  65&  5.25&    7.74&  8.82 &  -1.2& 1.08  &2.65\\
AGC208399    & 20.0 & 0.12& 24.6& 561 & 17.3&  31& 10.1 &    6.94&  8.04 &  -2.3& 1.10  &2.58\\
And IV       & 16.6 & 0.26& 25.6& 503 & 14.0&  90&  7.17&    7.44&  8.43 &  -1.9&  .99  &2.38\\
VCC0169      & 17.7 & 0.08& 25.9&2094 & 15.6&  24&  9.4 &    7.16&  8.04 &  -0.3&  .88  &2.02\\
KKH38        & 17.4 & 0.40& 26.5& 518 & 15.4&  58& 10.2 &    7.46&  8.18 &  -2.6&  .72  &1.60\\ 
LeoT         & 16.5 & 0.13& 26.6& -97 & 14.9&  15&  0.42&    4.94&  5.63 &   0.7&  .69  &1.47\\
AGC205268    & 17.4 & 0.13& 23.8&1001 & 15.9&  70& 10.4 &    7.38&  8.03 &   1.1&  .65  &1.37\\
KK195        & 17.1 & 0.27& 24.8& 345 & 15.5&  24&  5.22&    6.96&  7.56 &   0.9&  .60  &1.33\\
DDO143       & 16.0 & 0.08& 26.2& 616 & 14.6&  50&  9.1 &    7.80&  8.43 &  -0.5&  .63  &1.32\\
LV J0926+3343& 17.8 & 0.08& 25.9& 492 & 16.4&  47& 10.3 &    7.85&  8.37 &  -1.5&  .52  &1.32\\
KK144        & 16.5 & 0.09& 25.3& 449 & 15.1&  38&  6.2 &    7.27&  7.86 &  -0.9&  .59  &1.31\\
AGC208397    & 19.7 & 0.17& 24.3& 573 & 18.3&  33& 10.2 &    7.28&  7.83 &  -3.0&  .55  &1.23\\
LSBC D564-08 & 17.9 & 0.12& 25.7& 366 & 16.7&  49&  8.67&    7.02&  7.52 &  -0.3&  .50  &1.08\\
KDG192       & 16.6 & 0.06& 25.3& 544 & 15.5&  69&  7.4 &    7.38&  7.88 &  -0.1&  .50  &1.04\\
UGCA292      &ф 16.1 & 0.07& 24.7& 306 & 15.0&  27&  3.61&    6.73&  7.44 &  -0.6&  .71  &1.03\\
KDG215       & 16.9 & 0.09& 25.8& 362 & 15.8&  25&  4.83&    7.01&  7.50 &  -0.7&  .49  &1.01\\
\hline                                                                                           
Mean         & 17.2 & 0.19& 25.5& 557 & 15.5&  45&  7.40&    7.14&  7.84 &  -1.0&  .70 & 1.55\\

\hline
\end{tabular}
\end{table*}
\end{document}